# Reducing Surface Wetness Leads to Tropical Hydrological Cycle Regime Transition

Bowen Fan[1], Zhihong Tan[2], Tiffany A. Shaw[1], and Edwin S. Kite[1]

[1]Department of the Geophysical Sciences, University of Chicago, Chicago, IL, USA, [2]Program in Atmospheric and Oceanic Sciences, Princeton University, Princeton, NJ, USA

**Abstract** Earth's modern climate is characterized by wet, rainy deep tropics, however, paleoclimate and planetary science have revealed a wide range of hydrological cycle regimes connected to different external parameters. Here we investigate how surface wetness affects the tropical hydrological cycle. When surface wetness is decreased in an Earth-like general circulation model, the tropics remain wet but the transition from a rainy to a rain-free regime. The rain-free regime occurs when surface precipitation is suppressed as negative evaporation (surface condensation) balances moisture flux convergence. The regime transition is dominated by near-surface relative humidity changes in contrast to the hypothesis that relative humidity changes are small. We show near-surface relative humidity changes responsible for the regime transition are controlled by re-evaporation of stratiform precipitation near the lifting condensation level. Re-evaporation impacts the near-surface through vertical mixing. Our results reveal a new rain-free tropical hydrological cycle regime that goes beyond the wet/dry paradigm.

**Plain Language Summary** Paleoclimatology and planetary science have revealed a range of tropical climates, for example, the Earth's modern deep tropics are wet and rainy, whereas the tropics of Snowball Earth and Titan are drier. Previous work showed a range of external parameter affects the tropical hydrological cycle. Here we quantify how surface wetness alone affects the tropical hydrological cycle. In response to reduced surface wetness in an Earth-like general circulation model, the tropics remain wet but the transition from a rainy to a rain-free regime at the surface. The regime transition is controlled by the re-evaporation of rain near the cloud base. In the rain-free regime, tropical rainfall is re-evaporated aloft and impact near-surface through vertical mixing, increasing the relative humidity. The surface evaporation becomes negative (surface condensation occurs) because the near-surface air is wetter than the dry surface. Our results reveal a new tropical hydrological cycle regime that is wet but rain-free at the surface and shows wet/dry is not sufficient to characterize the tropics.

## 1. Introduction

A defining feature of Earth's modern climate is that the deep tropics are wet and rainy (Hartmann, 2016), where precipitation exceeds evaporation and there is moisture flux convergence by the large-scale circulation. However, paleoclimatology and planetary science have revealed a range of tropical climates. For example, the tropics in Snowball Earth simulations are dry, that is, evaporation exceeds precipitation (Abbot et al., 2013; Pierrehumbert, 2005). The hydrological cycle regime on Titan depends on the hydraulic conductivity $k$ of the surface. When $k$ is small or zero the tropics are in between wet and dry, i.e., precipitation equals evaporation (Liu & Schneider, 2016; Mitchell, 2008), whereas when $k$ is large the tropics are wet (Faulk et al., 2020). Understanding the full spectrum of tropical hydrological cycle regimes and how the tropics transition between regimes are important for characterizing the hydrological cycle in our solar system and also for the habitability of exoplanets (e.g., Abe et al., 2011; Kalidindi et al., 2018; Kodama et al., 2018, 2019; Leconte et al., 2013).

Previous work uncovered different tropical hydrological cycle regimes by examining different climates that involve different external parameter settings, e.g., surface albedo, surface moisture availability or wetness, and orbital parameters (Abbot et al., 2013; Faulk et al., 2020; Liu & Schneider, 2016; Lora et al., 2014; Pierrehumbert, 2005). Here we seek to isolate the role of surface wetness on the tropical hydrological cycle regime. We focus on surface wetness because we hypothesize decreasing surface wetness will expose a new tropical hydrological cycle regime transition (see below).

There are several ways to parameterize surface wetness. Here we follow Cronin and Chavas (2019), and parameterize surface wetness in the equation for evaporation via the parameter $\beta$:

$$E = \rho C_K L_v V_s (\beta q_s^* - q_1) \quad (1)$$

where $\rho$ is the density of near-surface air, $C_K$ is the exchange coefficient of moisture, $L_v$ is the latent heat of vaporization of water, $V_s$ is surface wind speed, $q_s^*$ is the saturation specific humidity at the surface, and $q_1$ is the near-surface specific humidity. Physically $\beta$ is the mole fraction of water for an ideal solution where there are no interactions between the solute and solvent. Thus, $\beta$ is a proxy for salinity or other exotic ocean compositions, that is, small $\beta$ corresponds to a surface with limited water and high salinity. In arid environments like Antarctic ponds, $\beta$ can be as small as 0.28 (Toner et al., 2017). Our formulation for surface wetness also follows Mitchell et al. (2006, and 2009). An alternative approach to parameterizing surface wetness is to put $\beta$ outside of the parentheses in 1, which then acts as an evaporation resistance and represents dry soil. For example, see the moisture availability parameter $M$ in Equation 12 of Tokano et al. (2001).

Here we seek to answer the question "How does surface wetness affect the tropical hydrological cycle regime?" and start with the simplest Earth-like set up with no seasonal cycle. To highlight the role of near-surface relative humidity, we rewrite Equation 1 as

$$E = \rho C_K L_v V_s q_s^* (\beta - \mathcal{H}) \quad (2)$$

where we define:

$$\mathcal{H} = q_1 / q_s^*(T_s) \quad (3)$$

which is quantitatively similar to the near-surface relative humidity (Table S1). Following Held and Soden (2006) and Schneider et al. (2010), if we assume changes in relative humidity are small as surface wetness $\beta$ decreases, then we hypothesize a deep tropical regime transition from rainy to rain-free at the surface. Namely, surface evaporation will become negative as $\beta$ decreases and assuming negative surface evaporation is balanced by moisture flux convergence, then the tropics are rain free at the surface (surface precipitation is approximately zero). According to our hypothesis, varying surface wetness alone does not lead to a wet to dry regime transition. Instead, it reveals a wet, rain-free regime where moisture condenses at the surface without surface rainfall. Our hypothesis assumes relative humidity changes are small, which has been verified in response to surface warming but must be tested for climate changes induced by decreased surface wetness. We test our hypothesis by decreasing surface wetness in an Earth-like general circulation model (GCM).

In the following sections, we begin by introducing the GCM simulations and moisture budget analysis (Section 2). We then present and discuss the tropical precipitation response to decreasing surface wetness and compare it to the hypothesis (Section 3). We conclude the study by summarizing our results and discuss directions for future work (Section 4).

## 2. Methods

### 2.1. GCM Simulations

We use the finite volume dynamical core of the GFDL-AM2 GCM with an aquaplanet configuration (Anderson et al., 2004). The simulations are configured as follows: diurnal cycle but no seasonal cycle (obliquity and eccentricity are zero); the mixed layer depth is 50 m with no ocean heat transport and no sea ice; greenhouse gas concentrations are $CO_2$ = 348 ppmv, $CH_4$ = 1650 ppmv, $N_2O$ = 306 ppbv, CFC-11 = 0, and CFC-12 = 0; ozone distribution is set as in Blackburn and Hoskins (2013). All simulations are run for 60 years with 10 years of spin up.

The GCM simulates two types of precipitation: Stratiform and convective. Stratiform precipitation falls from nimbostratus clouds, which are produced by the Tiedtke-Roststayn-Klein prognostic cloud scheme (Jakob & Klein, 1999; Rotstayn, 1997; Tiedtke, 1993). Convective precipitation falls from cumulus and cumulonimbus clouds, which are produced by the relaxed Arakawa-Schubert scheme (Moorthi & Suarez, 1992).

FAN ET AL. 2 of 9

In the stratiform scheme, the re-evaporation of rain is calculated in each atmospheric layer by integrating the diameter-dependent evaporation rate of a single raindrop over the Marshall-Palmer droplet size distribution (Marshall & Palmer, 1948). Re-evaporation of rain only happens if the relative humidity in the unsaturated part of the grid box $RH_{clr}$ is less than a critical value $RH_{evap}$.

In order to understand the importance of re-evaporation of rain for the tropical hydrological cycle, we set up mechanism denial experiments. Mechanism denial experiments involve disabling a physical effect in the model in order to test its importance. We disable the re-evaporation of rain in the stratiform scheme by setting $RH_{evap}$ to 0. We can also disable the re-evaporation in the relaxed Arakawa-Schubert scheme, but that does not cause a significant change in the precipitation.

### 2.2. Moisture Budget

We focus on the hydrological cycle in the deep tropics, which we define as a meridional average from 5°S to 5°N and encapsulates the width of the ITCZ (Table S2). We connect surface wetness to precipitation using the atmospheric moisture budget because surface wetness appears as an external control parameter. The atmospheric moisture budget is:

$$P = E - \nabla \cdot \vec{F}_q \quad (4)$$

where $P$ is surface precipitation, $\nabla \cdot \vec{F}_q$ is the column-integrated moisture flux divergence and $E$ is surface evaporation. The surface wetness ($\beta$) appears in surface evaporation (see Equation 1). We performed simulations with $\beta = 0$, 0.001, 0.003, 0.01, 0.03, 0.1, 0.3, 0.5, 0.7, and 1.0.

In order to identify regimes that are rainy or rain-free at the surface, we non-dimensionalize the atmospheric moisture budget by dividing it by the column-integrated moisture flux convergence, which is positive definite in the deep tropics across the range of $\beta$ values we consider, that is,

$$\hat{P} = \hat{E} + 1 \quad (5)$$

where $\hat{P} = -P / \nabla \cdot \vec{F}_q$ and $\hat{E} = -E / \nabla \cdot \vec{F}_q$. The rainy regime occurs when surface precipitation is large compared to surface evaporation ($\hat{P} \approx 1$ and $\hat{E} \approx 0$) whereas the rain-free regime occurs when surface precipitation goes to zero and surface evaporation is negative ($\hat{P} \approx 0$ and $\hat{E} \approx -1$). Note the rain-free regime corresponds to a tropical climate with a negligible precipitation efficiency (ratio of surface precipitation to net condensation, Langhans et al., 2015; Lutsko & Cronin, 2018; Narsey et al., 2019; O'Gorman, 2015).

Our hypothesis can be connected to the regimes defined by non-dimensional $\hat{E}$ and $\hat{P}$ as follows. Assuming we start from an Earth-like climate with values for $\rho$, $C_K$, $L_v$, $V_s$, $q_s^*$, and $\mathcal{H}$ in Equation 2 corresponding to $\beta = 1$, then we solve for $\beta$ such that $\hat{E} = -1$. The value corresponding to $\hat{E} = -1$ is $\beta = 0.65$. Thus our hypothesis will be verified if the rainy to rain-free regime transition occurs around this value.

In addition, we quantify our hypothesis that relative humidity changes are small by decomposing the evaporation change $\delta E$ as follows:

$$\delta E \approx \rho C_K L_v V_s [q_s^* \delta \beta - q_s^* \delta \mathcal{H} + (\beta - \mathcal{H})\delta q_s^* + \delta(\beta - \mathcal{H})\delta q_s^*] + \cdots \quad (6)$$

where $\delta E = E_{\beta_2} - E_{\beta_1}$ and $\beta_2$ is the adjacent $\beta$ value smaller than $\beta_1$. The right hand side represent the effect of decreasing surface wetness ($q_s^* \delta \beta$), near-surface relative humidity changes ($-q_s^* \delta \mathcal{H}$), saturation specific humidity changes [$(\beta - \mathcal{H})\delta q_s^*$] and finally the nonlinear effect of saturation specific humidity changes and deviations of $\beta$ from $\mathcal{H}$ [$\delta(\beta - \mathcal{H})\delta q_s^*$]. Other terms are negligible (Figure S1). If $\delta \mathcal{H} \approx 0$ then our hypothesis is confirmed.

## 3. Results

### 3.1. Response to Decreasing Surface Wetness

As surface wetness ($\beta$) decreases, the tropical hydrological cycle transitions smoothly from a rainy to a rain-free regime (Figure 1a). The rainy regime ($\hat{P} \approx 1, \hat{E} \approx 0$) corresponds to the hydrological cycle of modern

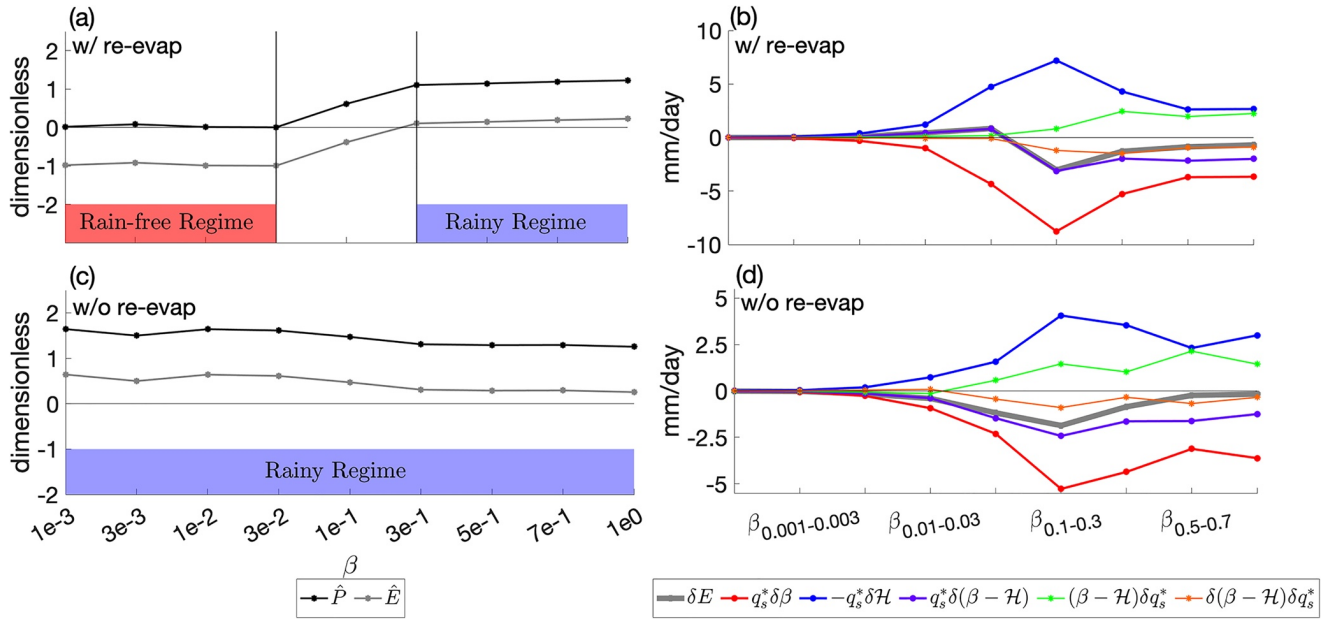

**Figure 1.** Time-, zonal- and tropical-mean (a) non-dimensional atmospheric moisture budget and (b) decomposition of surface evaporation following Equation 6 for simulations with re-evaporation (w/re-evap); (c), (d) are similar to (a), (b) but for simulations with re-evaporation disabled (w/o re-evap). Note the blue and red lines are re-scaled by a factor of 1/2 for better visualization.

Earth (Figure 1a) and occurs when $\beta \geq 0.3$. The rain-free regime ($\hat{P} \approx 0, \hat{E} \approx -1$) occurs when $\beta \leq 0.03$ (Figure 1a). The regime transition occurs for $0.03 < \beta < 0.3$, and does not exhibit hysteresis (Figure S2). It is also robust to the addition of a seasonal cycle (Figure S3).

While our hypothesis of a rainy to rain-free regime transition is verified (see Figure 1a) the regime transition does not occur around $\beta = 0.65$ as hypothesized. The decomposition of changes in evaporation also reveals that changes in relative humidity are not small compared to changes in $\beta$ in contrast to our hypothesis (compare red and blue lines in Figure 1b). In fact, the regime transition, which occurs between $0.03 < \beta < 0.3$, coincides with $q_s^* \delta \mathcal{H}$ dominating over $q_s^* \delta \beta$ (purple line, Figure 1b). This suggests that the relationship between $\beta$ and relative humidity $\mathcal{H}$ is important for the regime transition.

Indeed the transition from rainy to rain-free tropics defined by $\hat{E}$ and $\hat{P}$ can be connected to the relationship between $\beta$ and $\mathcal{H}$. For example, a necessary but not sufficient condition for the rainy regime is $\beta - \mathcal{H} \geq 0$ whereas the necessary but not sufficient condition for the rain-free regime is $\beta - \mathcal{H} \leq 0$ (thick black line, Figure 2). Given the failure of our hypothesis to account for relative humidity changes, we move on to understand what controls the change in relative humidity as $\beta$ decreases.

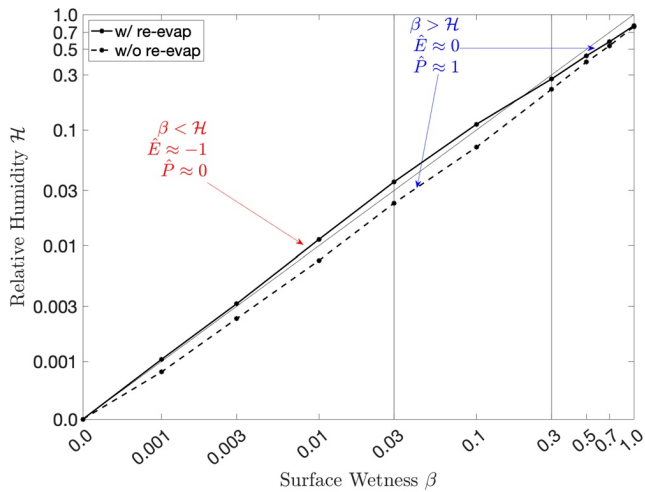

**Figure 2.** Time-, zonal- and tropical-mean near-surface (996 hPa) atmospheric relative humidity $\mathcal{H}$ as a function of surface wetness $\beta$ for simulations with (w/re-evap) and without (w/o re-evap) re-evaporation. The rainy regime occurs when $\beta \geq 0.3$ (w/re-evap) and with all values of beta (w/o re-evap), while the rain-free regime occurs when $\beta \leq 0.03$ (w/ re-evap). The thin black line from the bottom-left to the top-right is the one-to-one line. Vertical lines indicate the regime-transition region for simulations with re-evaporation.

### 3.2. Role of Re-Evaporation for the Rainy-to-Rain-Free Regime Transition

Near-surface relative humidity $\mathcal{H}$ is important for the rainy-to-rain-free regime transition of the tropical hydrological cycle. The relative humidity is affected by many different factors including re-evaporation within the atmosphere. Re-evaporation has been shown to have a large impact on the position of the ITCZ because re-evaporative cooling weakens the coupling between condensational heating and vertical motion (Bacmeister et al., 2006).

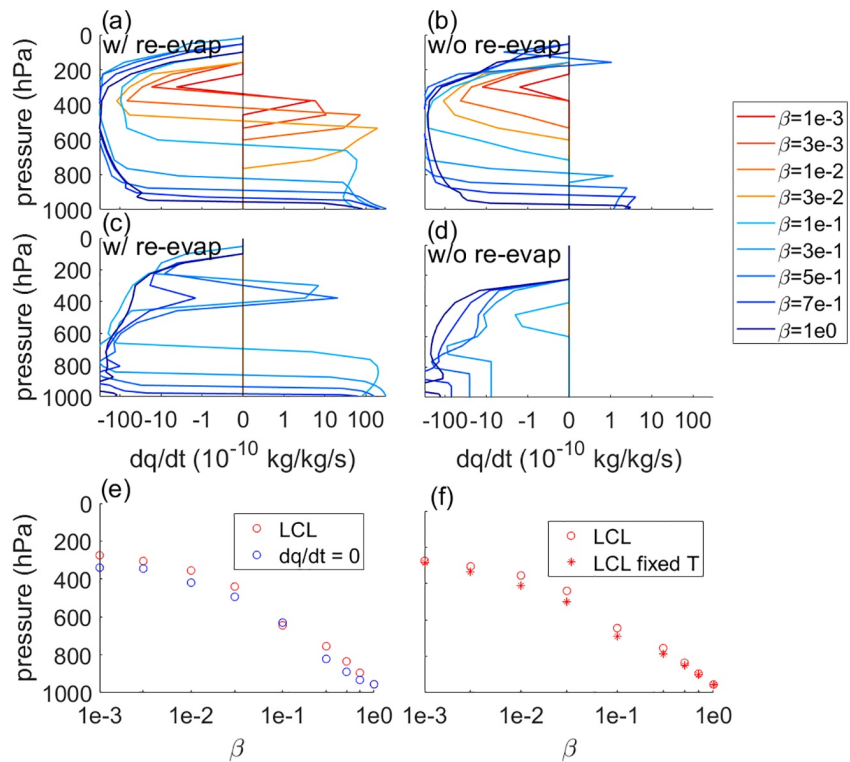

**Figure 3.** Moisture tendency from the stratiform precipitation scheme for simulations (a) with re-evaporation and (b) without re-evaporation and the convective precipitation scheme for simulations (c) with re-evaporation and (d) without re-evaporation. (e) Vertical pressure level where moistening tendency is zero with re-evaporation (blue circles) and lifting condensation level (LCL, red circles) predicted from Romps (2017) versus $\beta$ for simulations with re-evaporation. (f) Vertical pressure level of LCL versus $\beta$ using surface relative humidity and temperature (red circles) and with surface temperature fixed at $\beta = 1$ (red stars) for simulations with re-evaporation.

We quantify the impact of re-evaporation on the rainy to rain-free regime transition using mechanism denial experiments (see Section 2.1). When re-evaporation is disabled in the GCM simulations, there is no rainy to rain-free regime transition. The rainy regime occurs for all $\beta$ values (Figure 1c). Consistently, $q_s^*\delta(\beta - \mathcal{H}) \leq 0$ (purple line, Figure 1d) in the decomposition of evaporation changes, and $\beta - \mathcal{H} \geq 0$ (dashed black line, Figure 2) as $\beta$ decreases. Thus, near-surface relative humidity is strongly affected by re-evaporation.

### 3.2.1. Impact of Re-Evaporation From Stratiform Precipitation

Is re-evaporation associated with stratiform or deep convective precipitation? In the GCM simulations, both deep convective and stratiform precipitation are suppressed as $\beta$ decreases with re-evaporation enabled (Figure S4a). In the rain-free regime, the moistening tendency due to re-evaporation in the deep convection scheme is negligible compared to those in the stratiform scheme (compare Figures 3a to 3c). Disabling re-evaporation leads to enhanced stratiform precipitation but no impact on deep convective precipitation (Figure S4b). Consistently, the moistening tendency due to re-evaporation from stratiform precipitation dominates over that from the deep convection scheme in our model (compare Figures 3b to 3d).

Does re-evaporation from stratiform precipitation moisten the near-surface air via local (near-surface) or non-local (aloft) processes? The stratiform scheme output from the GCM suggests that re-evaporation moistens the air non-locally (aloft) as $\beta$ decreases (see $dq/dt > 0$ in Figure 3a) and generates a local maximum of specific humidity in the vertical near the Lifting Condensation Level (LCL, Figure S5). Moreover, when re-evaporation is disabled the moistening aloft does not occur (Figure 3b). Note the positive moisture tendency without re-evaporation in the lower atmosphere is associated with turbulent mixing of cloud droplets and the unsaturated environmental air, leading to cloud erosion and moistening of the environment (Figure S6).

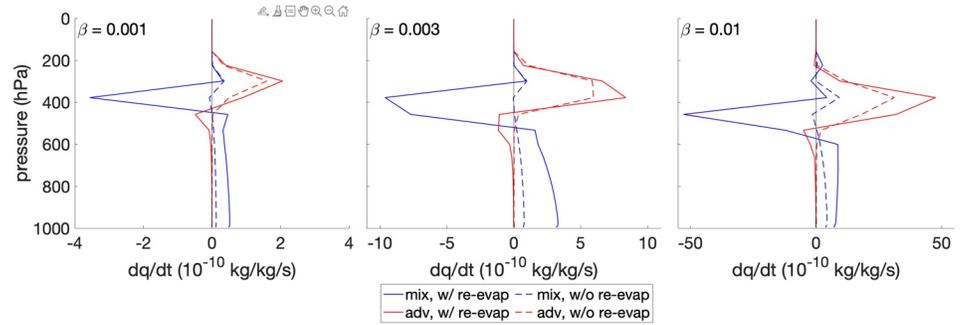

**Figure 4.** Rain-free regime moisture tendency from vertical mixing (blue lines) and vertical advection (red lines) for simulations with re-evaporation (solid) and without re-evaporation (dashed). When re-evaporation is enabled, vertical mixing transports moisture away (negative) from the lifting condensation level down to the near-surface (positive).

Why does re-evaporation occur aloft as $\beta$ decreases? The level where re-evaporation from stratiform precipitation moistens the atmosphere (blue circles, Figure 3e) follows the LCL (red circles, Figure 3e). The LCL is calculated using the surface relative humidity and temperature following Romps (2017). The LCL calculation does not depend significantly on surface temperature (compare red circles and stars, Figure 3f), which changes with $\beta$ (Figure S7). Thus, as surface relative humidity decreases following $\beta$, the near-surface air is further away from saturation and the LCL moves upward.

### 3.2.2. Role of Vertical Mixing

The re-evaporation near the LCL must impact the surface either through vertical advection by the large-scale circulation or vertical mixing by eddy diffusion. Note vertical mixing by eddy diffusion is the only parameterized mixing process in the GCM. Vertical mixing clearly moistens the atmosphere below the LCL in the rain-free regime (blue solid lines, Figure 4), whereas vertical advection by the large-scale circulation does not (red solid lines, Figure 4). The moistening tendency due to vertical advection is negligible in our simulations because the specific humidity $q$ is almost constant below the LCL (Figure S8). This is consistent with GCM simulations of Titan (Lora et al., 2015; Mitchell, 2008; Niemann et al., 2005) and Cloud Resolving Model (CRM) simulations of Earth's tropics with small surface wetness (Cronin & Chavas, 2019).

We can further confirm the dominance of vertical mixing by defining a vertical mixing time scale $\tau_{mix}$

$$\tau_{mix} = \frac{H^2}{\kappa} \quad (7)$$

where $H$ is the height of the LCL and $\kappa$ is extracted from the time-dependent model output. The vertical mixing time scale can be compared to the vertical advection timescale $\tau_{adv}$

$$\tau_{adv} = \frac{H_p}{\omega} \quad (8)$$

where $H_p$ is the pressure thickness of the LCL, that is, the difference between the surface pressure and the LCL, and $\omega$ is the mass-weighted average of pressure vertical velocity between the surface pressure and the LCL. In our simulations, $\tau_{mix}$ is much smaller than $\tau_{adv}$ (Table 1),

**Table 1**
*The Vertical Mixing Timescale $\tau_{Mix}$ and the Large-Scale Advection Timescale $\tau_{adv}$*

| $\beta$ | 0.001 | 0.003 | 0.01 |
|---|---|---|---|
| $H$(km) | 9.0 | 9.2 | 7.9 |
| $\kappa$ (m$^2$/s) | 1243 | 1132 | 892 |
| $H_p$ (hPa) | 725 | 695 | 625 |
| $\omega$ (Pa/s) | 0.143 | 0.146 | 0.131 |
| $\tau_{mix}$ (hour) | 18.1 | 20.8 | 19.4 |
| $\tau_{adv}$ (hour) | 140.8 | 132.2 | 136.8 |

*Note.* The vertical mixing timescale $\tau_{mix}$ is estimated by the height of LCL $H$ and diffusivity $\kappa$ (see Equation 7). The large-scale advection timescale $\tau_{adv}$ is estimated by pressure thickness of the LCL $H_p$, that is, the difference between the surface pressure and the LCL, and pressure velocity $\omega$. LCL, lifting condensation level.

confirming the dominance of vertical mixing for connecting re-evaporation near the LCL to surface relative humidity.

## 4. Summary and Discussion

A range of tropical hydrological cycle regimes have been documented across modern Earth, paleo- and planetary climates and are connected to a range of external parameters (surface albedo, surface wetness, orbital parameters, etc.). Here we isolated the impact of surface wetness on the tropical hydrological cycle. When surface wetness is decreased in an Earth-like GCM, the tropical hydrological cycle transitions from a regime that is wet with surface rainfall to a regime that is wet and rain free at the surface. The rain-free regime is associated with negligible surface rainfall and negative surface evaporation (surface condensation) but is still wet since precipitation is greater than evaporation.

The GCM results confirm our hypothesis of the emergence of a rain-free regime as surface wetness decreases, however, our hypothesis fails to capture when the regime transition occurs and the mechanism behind it. In particular, the simulations show the regime transition occurs between $0.03 < \beta < 0.3$ rather than $\beta = 0.65$ as hypothesized. Furthermore, near surface relative humidity changes are not small, and instead near-surface relative humidity changes are just as large as surface wetness changes. More specifically, a necessary but not sufficient condition for the rainy regime is $\beta \geq \mathcal{H}$ whereas the necessary but not sufficient condition for the rain-free regime is $\beta \leq \mathcal{H}$.

Mechanism denial experiments show that when re-evaporation of stratiform precipitation is disabled, $\beta \geq \mathcal{H}$ and there is no regime transition as surface wetness decreases. Thus re-evaporation affects the regime transition by impacting near-surface relative humidity. More specifically, re-evaporation generates a local maximum of specific humidity near the LCL then the moisture mixes vertically throughout the boundary layer. The role of vertical mixing is consistent with the mixing time scale being smaller than the timescale associated with vertical advection by the large-scale circulation.

Our aquaplanet results revealed a new rain-free regime that depends on re-evaporation of stratiform precipitation. We do not believe the importance of re-evaporation is an artifact of parameterized convection in our GCM. We find multiple similarities between our GCM results and the CRM results in Cronin and Chavas (2019). For example, in both cases reducing surface wetness leads to a deepening of the boundary layer and rising LCL, and similar changes of cloud fraction and precipitation flux (compare Figure S8 to their Figure 2).

Our results show characterizing the tropical hydrological cycle using only the relationship between precipitation and evaporation, that is, wet ($P > E$), dry ($P < E$) or in between ($P \approx E$), is not sufficient. It is also important to account for the impact of moisture flux convergence, that is, rainy ($\hat{P} \approx 1$) versus rain-free ($\hat{P} \approx 0$) regime. While wet and dry regimes have been documented in paleo- and planetary climates, the rain-free regime has not occurred in the geologic record to our knowledge. Snowball Earth is rainy and dry, whereas Titan (with large hydraulic conductivity) is rainy and wet with seasonal variations playing an important role in both cases. Nevertheless, the rain-free regime might apply to exoplanets (Abe et al., 2011; Kodama et al., 2018, 2019).

Finally, future work should focus on:

- How the regime transition depends on the exact formulation of surface wetness, that is, parameterization of salty ocean versus dry soil
- The connection between the tropics and other regions. For example, how negative surface evaporation in the deep tropics is sustained by positive surface evaporation and moisture flux divergence outside the tropics
- The connection between surface wetness and surface temperature
- The influence of other surfaces (albedo, heat capacity, topography) and orbital (obliquity, rotation rate) parameters, and the associated link to observed climates (Snowball Earth and Titan)

All of these things are important for improving our understanding of the factors affecting the tropical hydrological cycle.

## Data Availability Statement

The simulations in this study were completed with resources provided by the University of Chicago Research Computing Center. Data necessary to reproduce the figures in this study will be available via the University of Chicago's institutional repository Knowledge@UChicago (https://knowledge.uchicago.edu/record/2784).

**Acknowledgments**
B.F. and T.A.S. acknowledge support from National Science Foundation (AGS-1742944). The authors thank Tim Cronin and Jonathan Mitchell for helpful reviews that led to an improved manuscript.